# Modelling and Systematics of 1/2[521]$_v$ Quasiparticle Rotational Bands in N=99,101,103 Isotonic Chains

Pinky[1], Sushil Kumar[1*], Sukhjeet Singh[1] and A.K. Jain[1,2]

[1]*Department of Physics, Akal University Talwandi Sabo, Bathinda, Punjab-151302, India*
[2]*AINST, Amity University, Noida- 201313, India*
[*]*email: sushil.rathi179@gmail.com*

**Abstract**

We analyze the 1/2[521]$_v$ quasiparticle rotational bands along the N = 99, 101 and 103 isotonic chains using semi-empirical model in which first- and higher-order Coriolis terms are treated perturbatively. A dedicated Python routine optimizes five parameters bandhead energy, inertia A, decoupling *a*, and Coriolis coefficients *B* and *C* for each nucleus. The calculations match experimental level spacings to within ±5 keV. Staggering keeps a common phase across the three chains yet grows with spin and, on average, with both neutron and proton number. The inertia parameter rises smoothly with Z (e.g., A for N = 99: 10.6 to 12.4 keV), signaling a decrease in the moment of inertia as axial deformation weakens. Decoupling parameter *a* increases and then saturates for N = 99 and 103, indicating purer K = 1/2 structure at high Z; the N = 101 chain shows the opposite trend, pointing to stronger configuration mixing. The first-order Coriolis term *B* becomes increasingly negative with Z, whereas the higher-order term *C* shows chain-specific sign changes. The present work, explore the dynamics of various physical parameters associated with ν1/2[521] quasiparticle rotational bands and helps to understand observed signature effects.

## 1. Introduction

The present work focuses on modelling and systematics of the ν1/2[521] quasiparticle band along the isotone N = 99, 101 and 103 isotonic chains. Many low-lying ν1/2[521] bandheads manifest as K-forbidden isomers with half-lives from nanoseconds ($^{179}$Hf) to hours ($^{183}$Os). Long lifetimes arise because E2 or M1+E2 decay to the 3/2$^-$ or 5/2$^-$ band members involve ΔK=2, yielding high hindrance factors. In axially symmetric deformed odd-A nuclei, the evolution of single particle state is well described by Nilsson's model and each orbital is labeled by the asymptotic quantum numbers N,$n_z$,Λ,Ω. The label N is the principal oscillator quantum number, $n_z$ counts nodes along the symmetry axis, Λ is the projection of orbital angular momentum on that axis, and Ω is the projection of the total single-particle angular momentum on symmetric axis and it defines the band quantum number K = Ω and provide a fundamental ground for the observation of one-quasiparticle rotational bands. We employ a semi-empirical particle-plus-rotor model treating first- and higher-order Coriolis terms perturbatively to describe the ν1/2[521] rotational bands. A purpose-built Python code fits each nucleus by simultaneously optimizing five key parameters: the band-head energy $E_0$, the inertia coefficient $A$, the decoupling parameter $a$, and the first- ($B$) and second-order ($C$) Coriolis coefficients.

## 2. The Methodology

In odd-A nuclei, each intrinsic state gives rise to a rotational band which is a sequence of levels with spins I=K, K+1, K+2, etc. and energies follow the approximate rigid-rotor I(I+1) systematics. However, in the deviation of rotational energies is observed when the valence particle occupies low-Ω and high-j orbitals. This deviation is mainly due to Coriolis interaction between particle and rotational degree of freedom of the nucleus and can be understood in the framework of particle plus rotor model approach. However, in present work, we modelled the rotational spectra odd-A nuclei using perturbation method. The perturbation caused by Coriolis term to the rotational spectra can be expressed as [1]:

$$E(I) = E_0 + AX + BX^2 + C(-1)^{I+1/2}\left(I - \frac{1}{2}\right)\left(I + \frac{1}{2}\right)\left(I - \frac{3}{2}\right)$$

Here, $X(I) = I(I+1) + (-1)^{I+1/2} a\left(I + \frac{1}{2}\right)$ is the effect of first order Coriolis interaction and $a$ is the decoupling parameter which determine the magnitude of coupling strength between even-even core and valence particle. The parameters involved in above equation are the band-head energy $E_0$, the inertia coefficient $A$, the decoupling parameter $a$, and the first- ($B$) and second-order ($C$) Coriolis coefficients. The magnitude of coefficient in higher order terms is generally small but significant when the band is highly perturbed. In present work, we used above equation to modelled the rotational spectra of bands built over $1/2[521]_\nu$ observed in N = 99 [3-6], 101 [7-10] and 103 [11-15] isotonic chains.

## 3. Results and Discussions

In this paper, we adopted an empirical formula based on first and higher order perturbation caused by Coriolis interactions and modelled the energy spectra of one quasiparticle rotational bands based on $1/2[521]_\nu$ Nilsson's configuration in N = 99, 101, and 103 isotonic chains. As the band under discussion is a low-$\Omega$, high-j orbitals and hence due to Coriolis interaction between the angular momentum of the rotating core and the odd nucleon causes the splitting of two rotational branches one is called as favoured (lower in energy) and other is unfavored (higher in energy). This cause energy staggering in the rotational band and visualized by E(I)-E(I-1) vs Spin (I). The phase of the staggering tells us which signature is favoured, while the amplitude measures how strongly the particle is coupled to the rotation. A python script is designed to modelled the spectra of rotational bands under discussion in 13 different nuclides belong to N=99,101,103 isotonic chains. The results of present calculations are shown in Figure 1, Figure 2 and Figure 3, it is clear from this figures that, the model calculations and in excellent matching with experimental data. These figures show that, across the N=99,101,103 isotonic chains, the staggering keeps a common phase and its magnitude grows with spin and, on average with neutron number. A similar trend with proton number is also visible i.e. moving from lighter elements (Er, Yb, Hf) to heavier ones (W, Os), the energy staggering typically becomes larger. This is consistent with known shape evolution as the deformation is strongest near mid-shell and gradually decreases toward higher-Z in these chains which causes enhancing the K-mixing, and thus amplifies energy staggering. The optimized values of physical parameters ($E_0$, a, A, B and C) associated with present model is listed in Table 1.

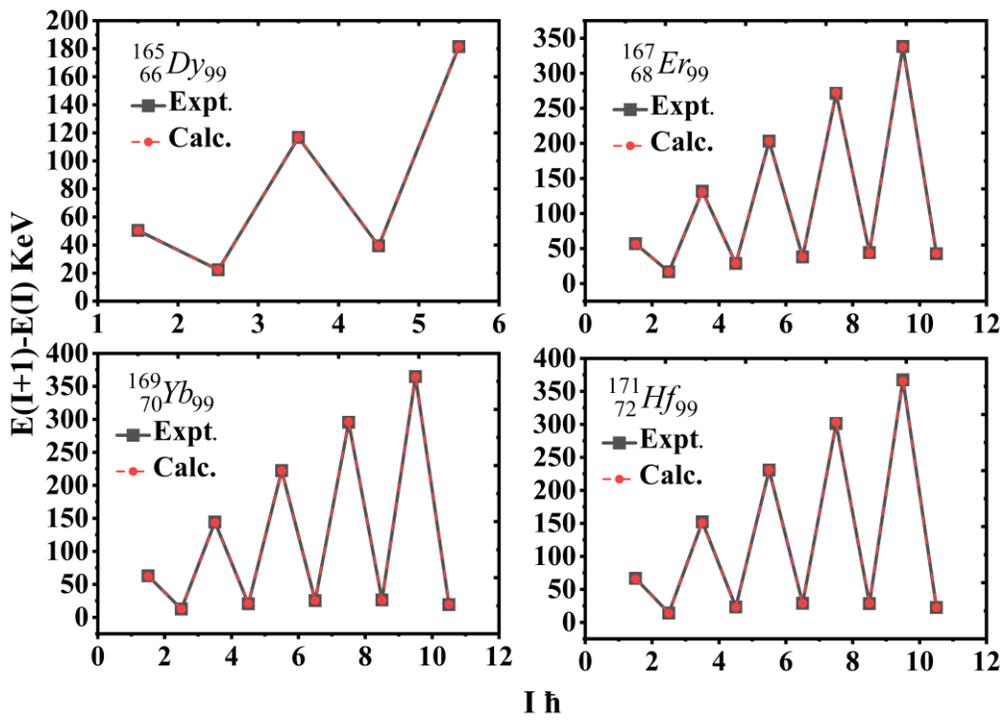

Figure 1: Comparison of calculated signature splitting with experimental data in N=99 isotonic chain.

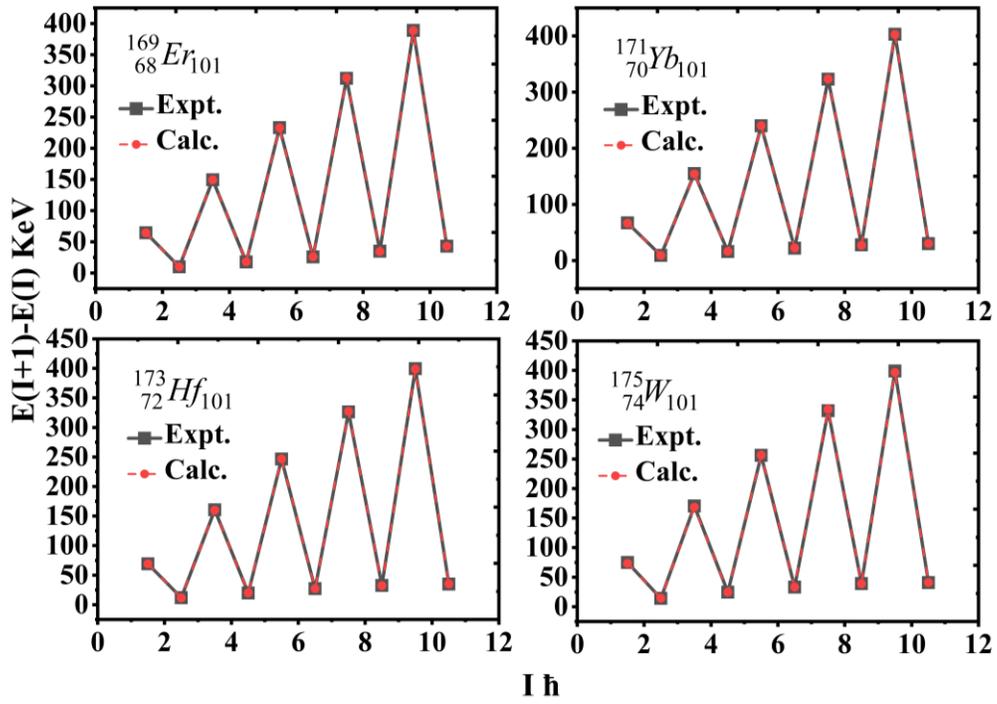

Figure 2: Same as figure 1 but for N=101 isotonic chain.

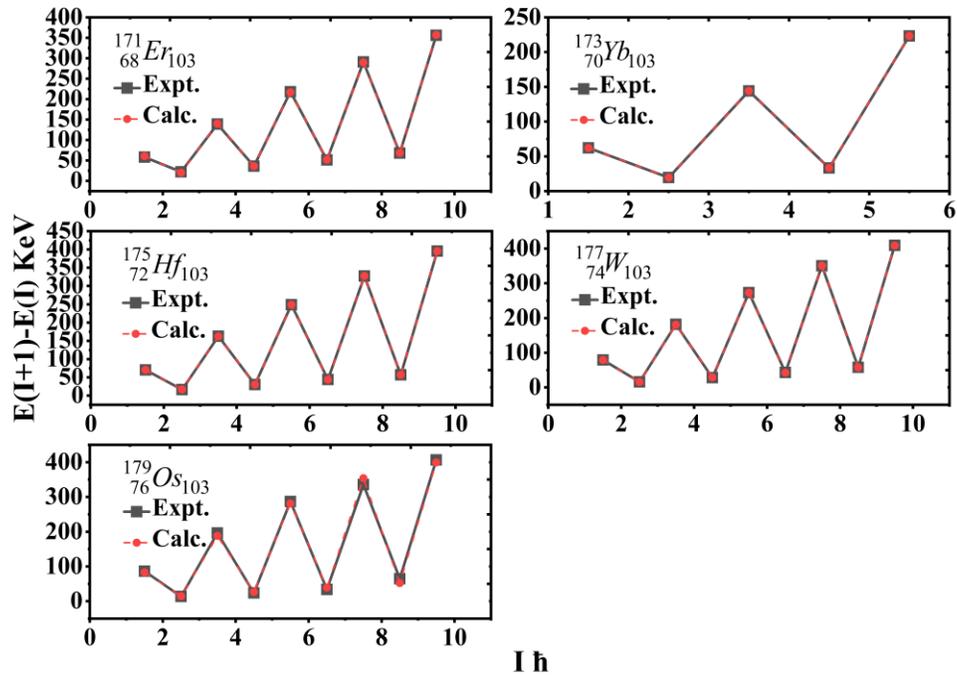

Figure 3: Same as figure 1 but for N=103 isotonic chain.

Table 1: The values of optimized parameters $E_0$, a, A, B and C obtained for $1/2[521]_\nu$ rotational bands observed in Er, Yb, Hf, W and Os nuclides in N=99,101 and 103 isotonic chains.

| N | Nuclide | E (keV) (expt) | E (keV) (theo) | $E_0$ (keV) | a | A (keV) | B (keV) | C (keV) |
|---|---|---|---|---|---|---|---|---|
| 99 | $^{165}$Dy | 108.2 | 108.2 | 106.3 | 0.58 | 10.66 | -0.0041 | 0.0025 |
| 99 | $^{167}$Er | 207.8 | 208.4 | 207.7 | 0.69 | 11.16 | -0.0062 | 0.0092 |
| 99 | $^{169}$Yb | 24.2 | 24.8 | 25.1 | 0.78 | 11.68 | -0.0082 | 0.0116 |
| 99 | $^{171}$Hf | 21.9 | 23.3 | 23.5 | 0.77 | 12.41 | -0.0116 | 0.0112 |
| 101 | $^{169}$Er | 0.0 | 0.2 | 1.1 | 0.83 | 11.74 | -0.0027 | -0.0015 |
| 101 | $^{171}$Yb | 0.0 | 0.5 | 1.6 | 0.84 | 12.02 | -0.0036 | 0.0028 |
| 101 | $^{173}$Hf | 0.0 | 0.4 | 1.2 | 0.82 | 12.76 | -0.0075 | 0.0035 |
| 101 | $^{175}$W | 0.0 | 1.7 | 2.2 | 0.79 | 13.76 | -0.0125 | 0.0034 |
| 103 | $^{171}$Er | 198.6 | 197.6 | 196.6 | 0.67 | 12.11 | -0.0040 | -0.0026 |
| 103 | $^{173}$Yb | 400.4 | 400.3 | 399.4 | 0.68 | 12.39 | -0.0076 | 0.0096 |
| 103 | $^{175}$Hf | 125.9 | 125.8 | 125.8 | 0.75 | 13.48 | -0.0078 | -0.0032 |
| 103 | $^{177}$W | 0.0 | 0.2 | 0.8 | 0.79 | 14.84 | -0.0130 | -0.0089 |
| 103 | $^{179}$Os | 0.0 | 4.7 | 5.6 | 0.81 | 15.41 | -0.0185 | -0.0086 |

In Figure 4 (a-c), we also presented the evolution of these parameters in the isotonic chain under discussions. The bandhead energy drops steeply from ≈856 keV at N=89 ($^{153}$Gd) to near the Fermi surface (≈24 keV) by N=99 ($^{169}$Yb) and becomes the ground state for most isotones with 101≤N≤103. This trend reflects the downward movement of the 1/2[521] orbital with increasing neutron number and the concomitant increase in deformation. The inertia parameter ($A=\hbar^2/2\mathcal{J}$): In all chains (N=99,101,103) it rises steadily with Z (e.g., N=99: 10.6→12.4 keV; N=101: 11.7→13.8 keV; N=103: 12.1→15.7 keV). Since $A=\hbar^2/2\mathcal{J}$, this implies a decrease of the effective moment of inertia with increasing Z which is consistent with the expected reduction of axial deformation along an isotonic chain.

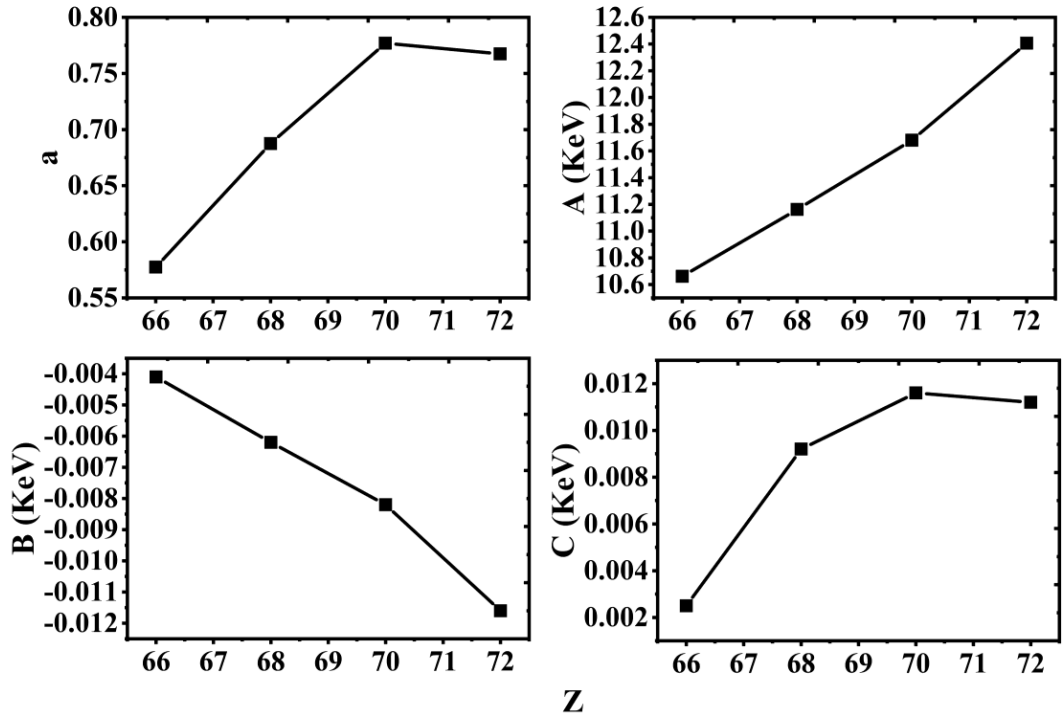

Figure 4 (a): Variation of physical parameters A, a, B, C in N=99 isotonic chain.

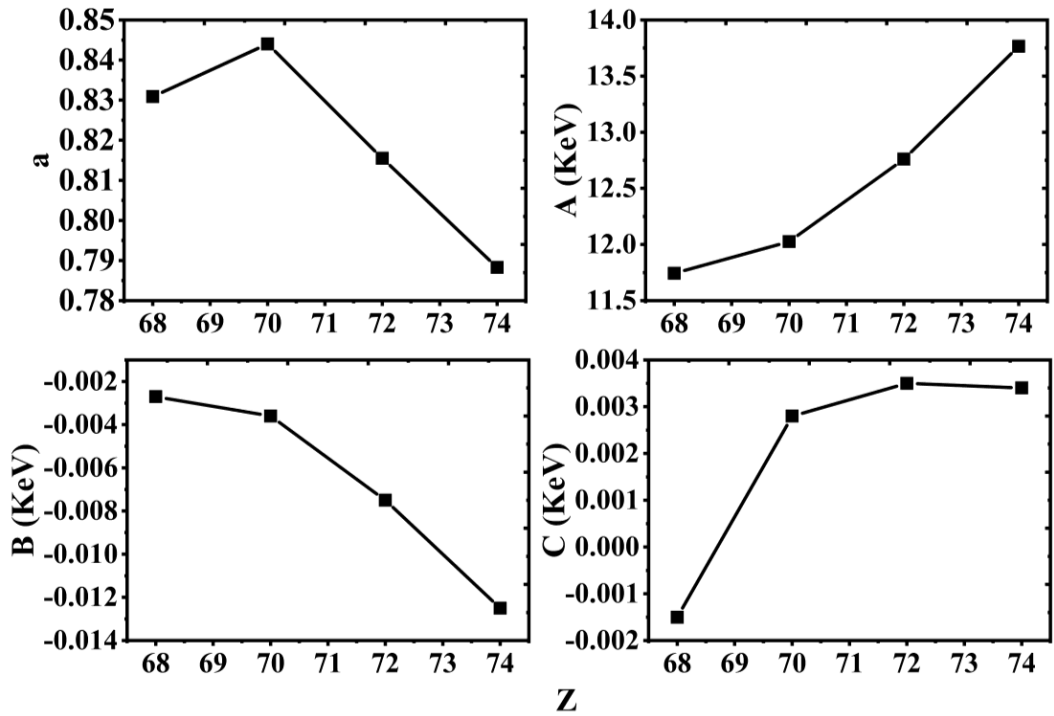

Figure 4 (b): Same as figure 4(a) but for N=101 isotonic chain.

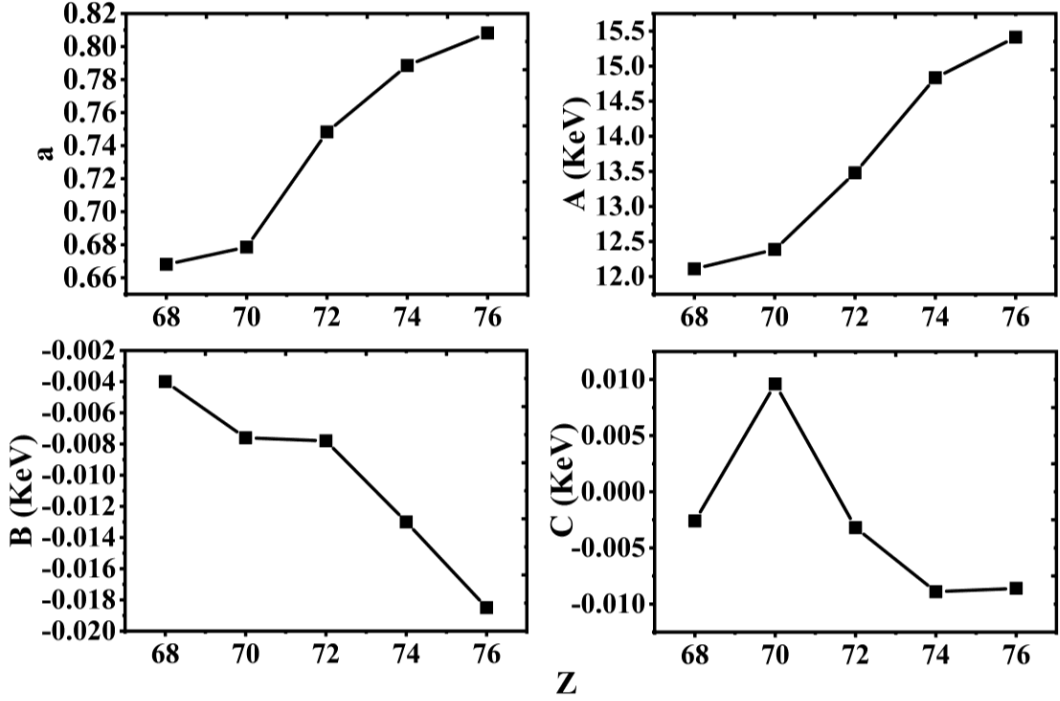

Figure 4 (c): Same as figure 4(a) but for N=103 isotonic chain.

The decoupling parameter $a$ in N=99 grows from ≈0.58 (Dy) to ≈0.78 (Yb) and then saturates (~0.77, Hf), indicating a trend toward a cleaner more decoupled $1/2[521]\nu$ configuration at higher Z. For N=101, it peaks near Z=70 (~0.845) and decreases to ~0.79 by Z=74 (W) pointing to stronger configuration mixing/K-mixing in the heavier elements of this chain and N=103 a monotonic increase i.e. (~0.67→0.81) again indicating progressive decoupling toward higher Z.

First-order Coriolis term $B$. The coefficient becomes more negative with Z in every chain (e.g. N=99: −0.004→ −0.012 keV; N=103: −0.004→ −0.019 keV), showing that the effective Coriolis influence strengthens as deformation decreases and K-mixing grows. The Higher-order Coriolis term $C$ is positive and increases up to Z=70 before saturating is consistent with mild higher-order effects in N=99 isotonic chain. N=101: $C$ evolves from slightly negative at Z=68 to positive near Z=72 and then become constant. However, N=103: $C$ shows a sign reversal (positive at Z=70, negative for Z≥72) and grows in magnitude toward W/Os. Overall picture. As Z increases in an isotonic chain the nuclei become less deformed and hence in $\mathcal{J}$ drops (so $A$ rises), and Coriolis effects intensify (larger $|B|$, non-zero $C$). The behaviour of $a$ distinguishes chains: N=99 and N=103 trend toward a purer 1/2[521]$_v$ band at high Z, whereas N=101 shows enhanced mixing for heavier elements. These systematic patterns provide a link between shape evolution and the strength and order of Coriolis coupling in K=1/2 bands.

## 4. Conclusions

The perturbative semi-empirical model accurately reproduces the 1/2[521]$_v$ rotational spectra in all N = 99, 101 and 103 isotones. The systematics of associated physical parameters (A, a, B and C) have been presented. Rising inertia parameters and increasingly negative B values confirm that axial deformation decreases and Coriolis mixing intensifies with proton number in a given isotone. Decoupling parameter *a* differentiates the chains: N = 99 and 103 trend toward more pure quasiparticle character at higher Z, whereas N = 101 exhibits enhanced K-mixing. The systematics of various physical parameters such as rotational and decoupling parameters for 1/2[521]$_v$ rotational bands across N≈99–103 shed light on the deformation and Coriolis interactions.